# Polaronic and Mott insulating phase of layered magnetic vanadium trihalide VCl$_3$

Dario Mastrippolito,[1, *] Luigi Camerano,[1] Hanna Swiatek,[2, 3] Břetislav Šmíd,[4] Tomasz Klimczuk,[2, 3] Luca Ottaviano,[1, 5] and Gianni Profeta[1, 5]

[1]*Department of Physical and Chemical Sciences,
University of L'Aquila, Via Vetoio, 67100 L'Aquila, Italy*
[2]*Faculty of Applied Physics and Mathematics Gdansk University of Technology, Gdansk, Poland*
[3]*Advanced Materials Center, Gdansk University of Technology, Gdansk, Poland*
[4]*Charles University, Faculty of Mathematics and Physics,
Department of Surface and Plasma Science, V Holešovičkách 2, 180 00 Prague 8, Czech Republic*
[5]*CNR-SPIN L'Aquila, Via Vetoio, 67100 L'Aquila, Italy*

Two-dimensional (2D) van der Waals (vdW) magnetic 3$d$-transition metal trihalides are a new class of functional materials showing exotic physical properties useful for spintronic and memory storage applications. In this article, we report the synthesis and electromagnetic characterization of single-crystalline vanadium trichloride, VCl$_3$, a novel 2D layered vdW Mott insulator, which has a rhombohedral structure (R$\bar{3}$, No. 148) at room temperature. VCl$_3$ undergoes a structural phase transition at 103 K and a subsequent antiferromagnetic transition at 21.8 K. Combining core levels and valence bands x-ray photoemission spectroscopy (XPS) with first-principles density functional theory (DFT) calculations, we demonstrate the Mott Hubbard insulating nature of VCl$_3$ and the existence of electron small 2D magnetic polarons localized on V atom sites by V-Cl bond relaxation. The polarons strongly affect the electromagnetic properties of VCl$_3$ promoting the occupation of dispersion-less spin-polarized V-3d $a_{1g}$ states and band inversion with $e'_g$ states. Within the polaronic scenario, it is possible to reconcile different experimental evidences on vanadium trihalides, suggesting that also VI$_3$ hosts polarons. Our results highlight the complex physical behavior of this class of crystals determined by charge trapping, lattice distorsions, correlation effects, mixed valence states, and magnetic states.

## I. INTRODUCTION

Layered van der Waals (vdW) 3$d$-transition metal trihalides (MX$_3$; M = Cr, V; X = I, Br, Cl) crystals are nowadays considered for integration into spintronic and memory storage applications at the nanoscale due to their intrinsic two-dimensional (2D) magnetism and exotic physical properties [1, 2]. Although this class of compounds has particularly rich physics, currently, the research interest is moving toward the vanadium trihalides (VX$_3$) group, as they show higher magnetic transition temperatures than their Cr-based counterparts [3, 4]. Vanadium tri-iodide, VI$_3$, was recently studied by different experimental and theoretical approaches to disclosing the intricate nature of its electronic ground state, which is determined by the interplay between crystal field effects, spin-orbit coupling, and strong correlations [5–7]. Photoemission experiments [6, 7] revealed VI$_3$ as a highly-correlated Mott insulator with low-dispersing Hubbard bands that fall into the valence bands (VB), pointing out to an electronic ground state with occupied $a_{1g}$ and half-filled $e'_\pm$ doublet states [7]. However, until now, there has been no uniform consensus on the nature of its electronic structure nor on the other vanadium halides compounds. It should be to understand to which degree the members of the VX$_3$ family share the same physics or VI$_3$ represents a unique example. In this article, we report the synthesis and characterization of a novel compound belonging to the vanadium trihalides class, namely VCl$_3$, accompanied by a thorough theoretical investigation via first-principles Density Functional Theory (DFT) calculations in combination with photoemission experiments. Our results also provide a theoretical interpretation that explains contrasting experimental and theoretical evidence of VI$_3$.

VCl$_3$ is predicted to be attractive for several applications since it has rich physics hosting quantum and valley Hall effects, high magnetoresistance, and skyrmions [8–11]. However, to date and to the best of our knowledge, only VI$_3$ and VBr$_3$ were successfully grown in the form of macroscopic single crystals [4, 12], allowing for a complete physical characterization. On the other hand, VCl$_3$ has not yet been studied in its single-crystalline form and its electronic properties are still experimentally unexplored. VCl$_3$ has been predicted to be a metal [9, 10, 13–15], in contrast to the insulating behavior of the isostructural VI$_3$, as well as a Mott insulator [16, 17]. For these reasons, we have grown macroscopic single crystals of VCl$_3$ and investigated them by means of X-ray photoemission spectroscopy (XPS) to access electronic core levels and VB simultaneously and clarify its electronic nature. At variance with VI$_3$ and VBr$_3$, which are usually grown by chemical vapor transport technique [4, 12], we successfully grew single crystalline VCl$_3$ by the less common self-selecting vapor growth (SSVG) technique [18]. Details on the growth procedure are provided in the dedicated section in the Appendix.

* email: dario.mastrippolito@graduate.univaq.it

The SSVG technique is usually used for the growth of II-VI and IV-VI semiconducting crystals, but more recently it also found application in the growth of some transition metal halides [19, 20].

## II. RESULTS AND DISCUSSION

We start our discussion by reporting the structural characterization of the as-grown crystals, which have a 3.0 ± 0.2 Cl/V atomic ratio double confirmed by energy-dispersive X-ray (EDX) analysis (see Fig. S1 [21]) and XPS. In its single-crystalline form, $VCl_3$ is a layered material having weak interlayer vdW interaction between each layer, whose thickness is 3.316 Å (Fig. 1a) and is formed by a honeycomb lattice of V atoms bonded to two Cl sublayers. Vanadium atoms are octahedrally coordinated with Cl atoms with a bond length of 2.415 Å. As visible from the scanning electron microscopy (SEM) image of Fig. 1b, the as-grown $VCl_3$ crystals have a form of plate-like hexagons with a size of 1.5×1.5 mm$^2$ and a thickness of less than 0.1 mm. The crystals have a layered lamellar stacking that is easily exfoliable down to the 2D phase. In Fig. 1c, we report the room-temperature X-ray diffraction (XRD) pattern of a single-crystalline $VCl_3$. Only *00l* reflections are detected endorsing the layered phase of $VCl_3$. For the LeBail profile refinement, we applied the rhombohedral $BiI_3$-type structure (R$\bar{3}$, No. 148), which gives a layer spacing $d_{00l}$ = 5.83 Å in good agreement with the one reported (5.78 Å) in Ref. [22]. The temperature-dependent heat capacity measurements show that, upon cooling, $VCl_3$ undergoes a structural phase transition at T$^\star$ = 103 K and an antiferromagnetic transition at T$_N$ = 21.8 K (Fig. 1d). Those temperatures are consistent with the previously reported for polycrystalline $VCl_3$ [23, 24].

Once the structure and the correct synthesis of the crystals have been ascertained, *in-situ* mechanically-exfoliated thin $VCl_3$ flakes were investigated by XPS (Fig. 2) for the analysis of the core levels (V-2p and Cl-2p) and VB. The shape of the Cl-2p core level spectrum is well described by a single Voigt doublet in contrast to the $CrCl_3$ case, where a Cl-vacancy-related component is clearly detected in Ref. [25]. The Cl-2p doublet is centered at a binding energy (BE) of 198.1 eV, in agreement with the literature on $VCl_3$ powder [22], and has a spin-orbit (SO) splitting of 1.6 eV. The V-2p core level has its main V-2p$_{3/2}$ component at a BE of 515.3 eV, as for the $VCl_3$ powder [22], and a SO splitting of 7.5 eV. The V-2p is deconvoluted in three Voigt doublets revealing a multiplet structure, that is ubiquitous in the MX$_3$ family [25–27]. The most dominant component at a BE of 515.3 eV, accounting for 81% of the total V-2p spectral area, is assigned to $V^{3+}$, fully consistent with the octahedral coordination of the $V^{3+}$ with Cl$^-$ ions. The second component at lower BE, 514.3 eV, weighs 15% of the total V-2p area and is assigned to $V^{2+}$, similar to what was observed in $VI_3$ [6, 28]. The remaining 4% of the V-2p doublet at a BE of 516.8 eV, can be assigned to $V^{5+/4+}$ states [28, 29]. Oxidation of the thin $VCl_3$ flakes is excluded by the analysis of the nearby O-1s core level, which is composed of one peak (centered at a BE of 531.9 eV) assigned to molecular $O_2$ and/or organics [29, 30] reasonably originating from the underlying/surrounding carbon tape used as substrate (see Appendix). Indeed, any O-1s signal from vanadium oxides is expected at a much lower BE around 530 eV [28, 29]. The VB structure, reported in Fig. 2c, shows an evident insulating phase. Indeed, the measured VB extends from 8 eV to about 3 eV BE, where we observe a net VB edge. Two main broad features are observed at 6 eV and 4 eV and a clearly resolved and localized peak is detected just above the VB at a BE of 1.7 eV.

To understand the nature of the measured $VCl_3$ electronic states, we carried out first-principles DFT+U calculations (see Appendix for computational details). Both pristine $VCl_3$ monolayer and bulk cases were simulated but no substantial differences are found between the 2D and bulk phases (see Fig. S2 [21], S3 [21], and S4 [**?** ]) and including spin-orbit coupling (see Fig. S5 [21]) which is small and does not affect the resolution of the bands as for the other chlorine-based metal trihalide $CrCl_3$ [31]. The density of states (DOS) with and without SOC are reported for comparison in Fig. S6 [21]. We find a ferromagnetic Mott insulating ground state for the $VCl_3$ monolayer, as for the other Cl-based metal trihalide $CrCl_3$ [25, 32], whose VB states mainly originate from the hybridization between V-3d and Cl-3p states. The valence bands maximum (VBM) is constituted only of spin-up polarized states at the M point of the Brillouin zone resulting in an indirect bandgap of 1.8 eV (Fig. 3). However, the calculations of pristine $VCl_3$ are inconsistent with the reported XPS results. Firstly, the predicted charge state of V atoms is $V^{3+}$, incompatible with the observed $V^{2+}$ component in the V-2p core levels. Secondly, as evident in Fig. 2c, the DOS does not reproduce all features that emerge from the VB photoemission analysis. In fact, despite a satisfactory overall agreement in the BE region between 8 eV and 3 eV, the DOS does not predict any electronic states up to about 3 eV above the VBM, contrary to the experimental evidence. A critical inspection of the literature reveals that, indeed, these states were already detected also in other related systems, like $VI_3$, even if not completely noticed or discussed but assigned to some spurious effects [6, 7]. We think that their origin is still unclear and deserves a more careful theoretical explanation. As it will be demonstrated soon below, we attribute these states to the formation of polarons.

Polarons have been observed in several transition metal-based materials [33] and perovskite metal halides

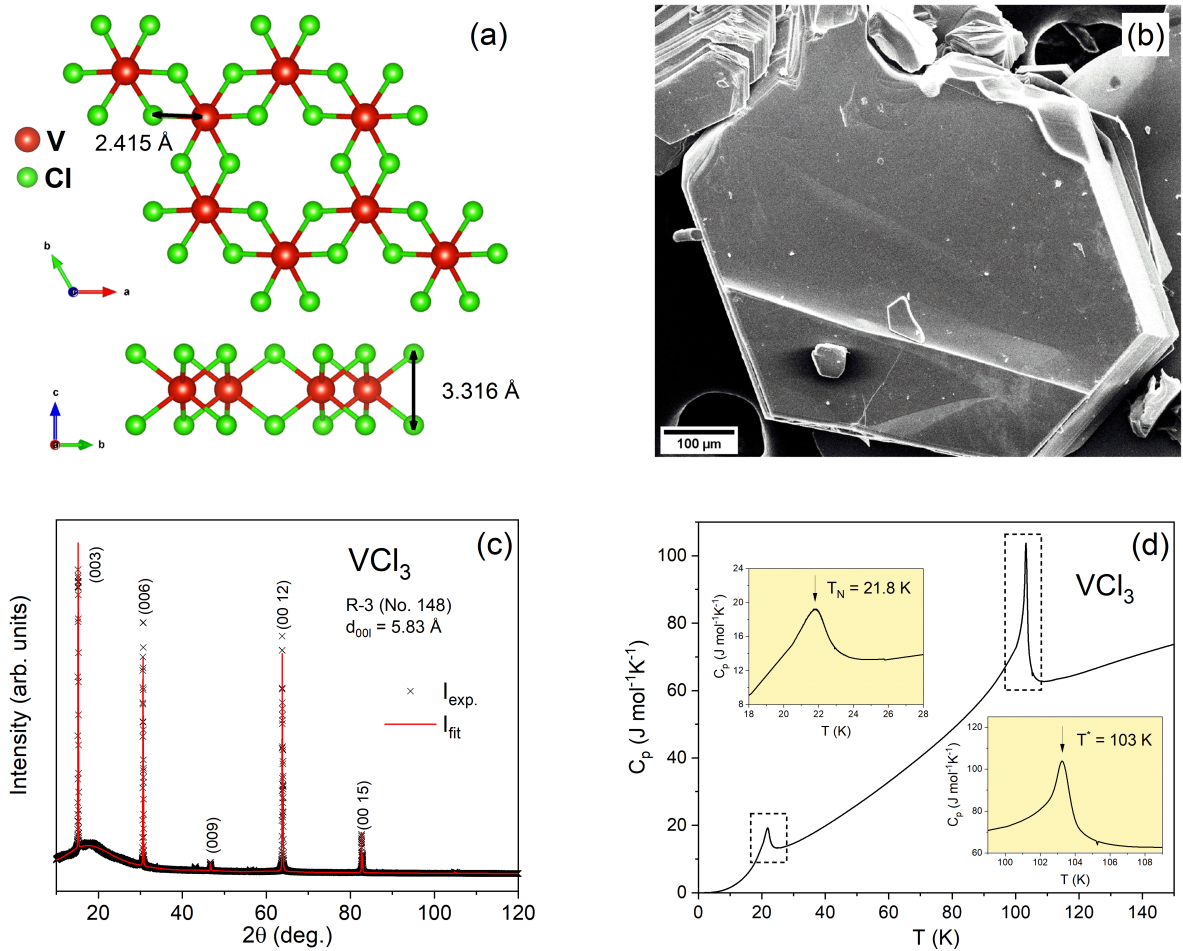

FIG. 1. (a) Top and side views of VCl$_3$ monolayer crystal structure. V and Cl atoms are represented in red and green spheres, respectively. (b) SEM image of VCl$_3$ flake-like crystals. (c) Room-temperature XRD experimental (black crosses) and R$\overline{3}$-model fitted (red line) spectra of VCl$_3$ crystal. (d) Temperature-dependent heat capacity of VCl$_3$ crystal. Insets: the antiferrogmanet phase transition at 21.8 K and the structural phase transition at 103 K.

[34], but rarely in the 2D phase [35]. However, very recently, evidences of polaronic states were founded in layered chromium trihalides [36–39]. Polarons change radically the electronic properties, spin transport, and also magnetic phases. In extremely correlated Mott insulating ionic systems, like VCl$_3$, even small perturbations in the electronic structure can lead, possibly, to the formation of polarons, which can be observed by XPS experiments and understood from the core levels and VB analysis [40–42]. In this regard, the observed V$^{2+}$ core-level component leans towards the presence of an excess of electrons, whose origin could be ascribed to photo-excited core electrons [43], which remain trapped in the material, or can be related to the presence of unstructured defects [33]. In order to validate this hypothesis we simulated a charged 2×2 supercell with one extra electron corresponding to 0.2 % n-type doping that quantitatively accounts for the observed amount of V$^{2+}$ signal. We find that the electron, driven by lattice distortions, localizes onto one of the eight V atom sites in the supercell, breaking the symmetry of the lattice and displacing the Cl atoms around the newly formed V$^{2+}$ atom. By energy minimization calculations, we find that the formed electron polaron is energetically favored by 360 meV with respect to the corresponding symmetric electron-doped configuration in which all the vanadium atoms are equivalent and the electron equally delocalizes over the eight vanadium atoms. The trapped electron forms a small polaron in the relaxed VCl$_6$ octahedron structure resulting in a V-Cl bond length of 2.503 Å, enhanced by 4% with respect to the symmetric case (Fig. 4c and S6). The structural perturbation induced by the excess electron is strongly localized around V$^{2+}$ atom and extends up only to the second nearest neighbor VCl$_6$ octahedrons surrounding the polaron. In the polaronic phase, some of the V-3d states, originating from the displaced V site, undergo a band crossing from the conduction to the valence bands, being occupied and localized within the band gap of VCl$_3$ (Fig. 4a). It is interesting to note how the not-localized electronic

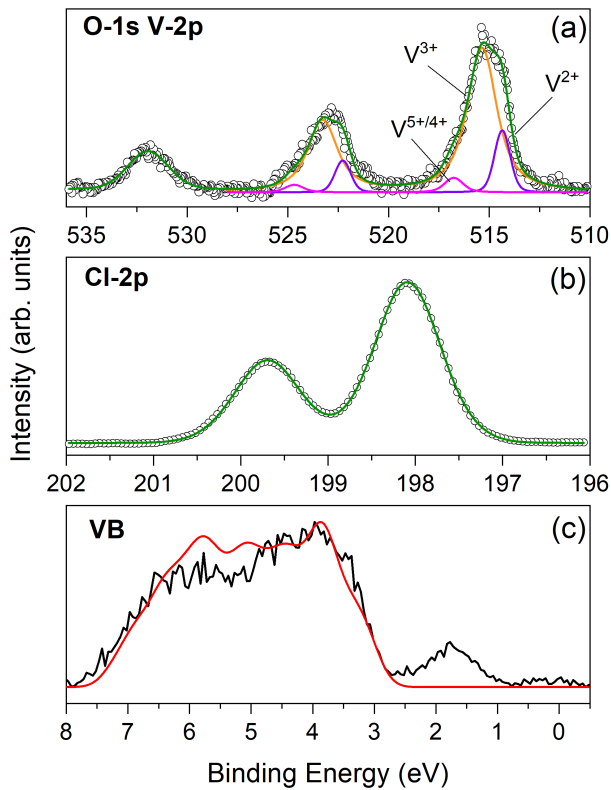

FIG. 2. XPS core levels and valence bands spectra of as-exfoliated VCl$_3$ flakes. The raw data (black), the cumulative fits (dark green), and the relative deconvolution (colored) are shown. (a) O-1s and V-2p core levels deconvolution. (b) Cl-2p core level deconvolution. (c) The experimental valence bands spectrum of VCl$_3$ flake (black) is directly compared to the DOS of pristine VCl$_3$ monolayer (red). The DOS is convoluted with a Gaussian function ($\sigma = 0.5$ eV) to account for the experimental broadening.

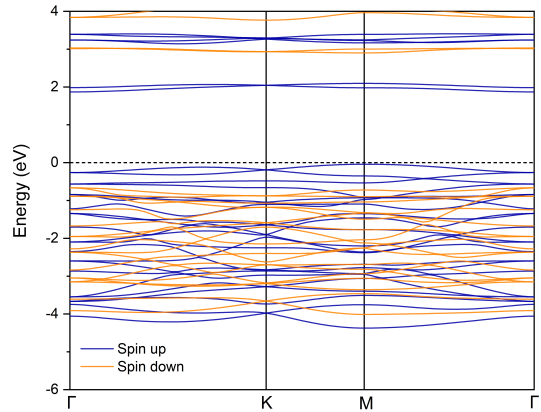

FIG. 3. Spin-resolved band structures of pristine VCl$_3$ monolayer

configuration results in a metallic phase (see Fig. S6a [21]) and, thus, excludes it as a candidate to explain the photoemission spectra. As clearly visible in Fig. 4b, the direct comparison between the calculated DOS, obtained by the polaronic mechanism, and the experimental VB shows that now the experimental spectrum is fully explained by the presence of electronic polaron localized on a single V site. Specifically, the shape of the VB spectrum and the presence of the localized peak above the VB are predicted by our calculations with excellent accuracy both in energy position and relative intensity. Notably, the localized peak detected in the experimental VB is a direct fingerprint of the polaronic V-3d state. The presence of polarons has important implications for electronic occupations (Fig. 4d). Indeed, in the undoped phase, the trigonal crystal field splitting of V$^{3+}$ atom forces the occupation of the two-fold $e'_g$ manifold in its ground state, while the $a_{1g}$ state is unoccupied [44]. On the contrary, in the polaronic V$^{2+}$ site, both $e'_g$ and $a_{1g}$ states are occupied, with inverted energy position. The $a_{1g}$ state has a strong $d_{z^2-r^2}$ orientation and, thus, it produces a non-negligible out-of-plane character of V-3d orbitals (Fig. 4e), which can profoundly alter the magnetic properties of VCl$_3$. In addition, the value of the magnetic moment increases to 2.6 $\mu_B$ for V$^{2+}$ with respect to the undoped V$^{3+}$ case which has a value of 1.9 $\mu_B$. Besides, it is worth examining whether the polaronic state satisfies the criteria for its existence in the 2D phase, as proved in a recent article by Sio *et al.* in Eq. 9 of Ref. [35]. For this reason, we calculated the static ionic dielectric constant of VCl$_3$ ($\epsilon_{ion} = 0.98$) and derived the effective mass of the polaronic band ($m_* \simeq 3.7\ m_e$). The values fully meet the existence requirements identifying the VCl$_3$ electron small polaron as a real 2D polaron.

Our discovery of magnetic polaron is not necessarily peculiar to the VCl$_3$ crystal but should be general and could exist also in other magnetic vdW materials hosting multiple valence atoms like vanadium and, in particular, in other VX$_3$ compounds, such as VI$_3$. As a confirmation of this, we will show below how the polaron formation naturally explains many recent (apparently contradicting) experiments on photoemission and X-ray magnetic circular dichroism (XMCD) [6, 7, 28, 45]. Indeed, in Refs. [6, 7, 45], it is claimed that VI$_3$ has occupied $a_{1g}$ state. However, magnetic phases in VI$_3$ obtained by forcing the occupation of the $a_{1g}$ orbital [5, 45–47], which can explain the large orbital moment observed in Ref. [45], fail to predict the electronic valence band of VI$_3$ and its sharp peak around 1 eV below the Fermi level (see Refs. [6, 7, 28] and Fig 5). To understand the origin of this anomalous sharp peak, in Ref. [6], the authors provide a tentative explanation proposing that the $a_{1g}$ would be occupied by external electron doping (probably band bending, etc.), which will fill the $a_{1g}$ state rigidly shifting and raising the Fermi level. Apart from the exceptionally high electron concentration needed to completely fill one band, which is on the order of $10^{12}$ e/cm$^2$, this interpretation is dis-

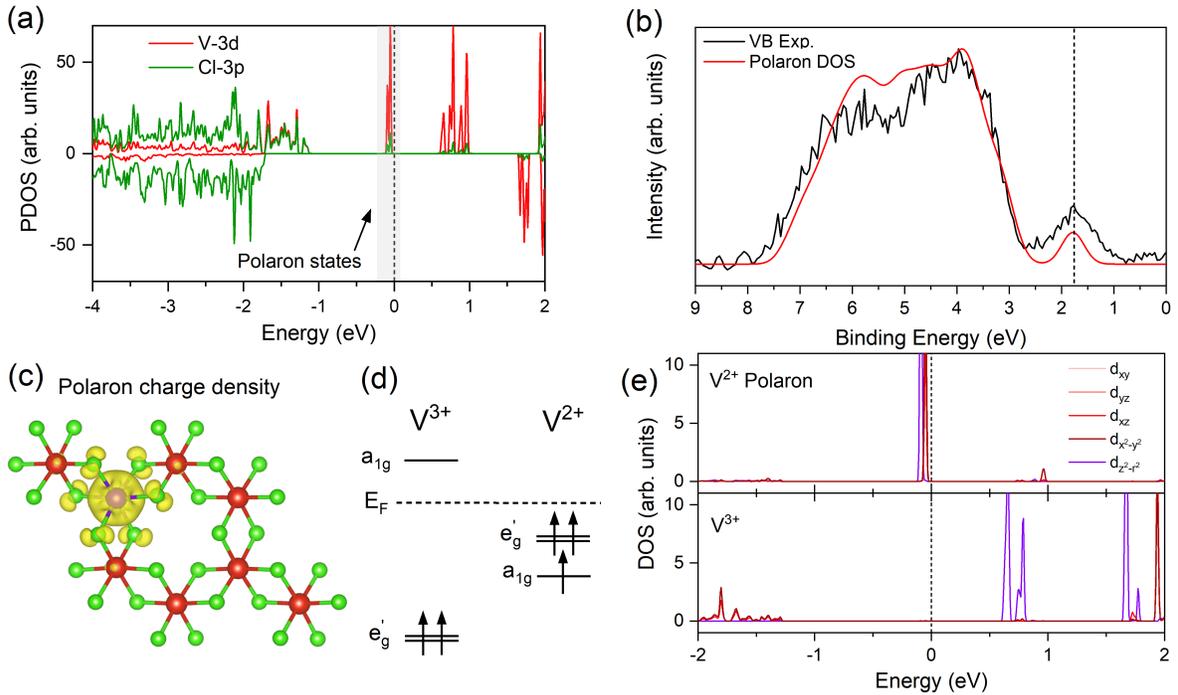

FIG. 4. Electron small 2D polarons of $VCl_3$. (a) Spin-resolved partial DOS of polaronic 2×2 supercell of $VCl_3$. Polaronic states are highlighted and contributed mainly by V-3d states. (b) The experimental valence bands spectrum of $VCl_3$ flake (black) is directly compared to the DOS of polaronic monolayer configurations (violet), convoluted with a Gaussian function ($\sigma = 0.5$ eV) to account for the experimental broadening. (c) Polaron charge density on the $VCl_3$ 2×2 supercell. V and Cl atoms are represented in red and green spheres, respectively. (d) Electrons occupation for $V^{3+}$ and $V^{2+}$ polaronic configurations. (e) Spatially-projected DOS of V-3d states for $V^{2+}$ polaronic atom and $V^{3+}$ atom.

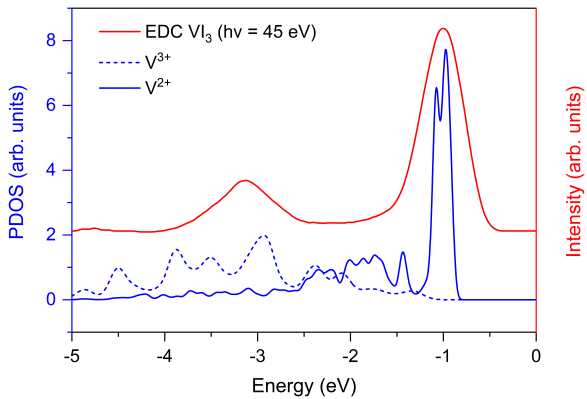

FIG. 5. Projected DOS for V-3d states for $V^{2+}$ polaronic atom (blue) and $V^{3+}$ atom (dashed blue) of $VI_3$ monolayer are compared with a $VI_3$ photoemission spectrum in resonance with V-3d states (red), which is adapted from Fig. 4a of Ref. [6].

proved by the experiments of Bergner *et al.* [7] in which direct n-type doping on $VI_3$ was induced via alkaline deposition. Unexpectedly and at odds with respect to the interpretation of De Vita *et al.* [6], the authors did not observe a (rigid) band shift or relevant modifications in the VB caused by the electron doping, but only a progressive increase of the spectral weight of the localized V-3d states as a function of deposition time. Conversely, in our polaronic framework, all these experimental occurrences would be naturally explained by the formation of electron polarons. Indeed, we simulated the DOS of the $VI_3$ monolayer with the same procedure. In Fig. 5, we explicitly compare the $VI_3$ photoemission spectrum in resonance with V-3d states (adapted from Fig. 4a of Ref. [6]) with our projected DOS on different vanadium sites within our polaronic hypothesis. The peak near -1 eV can be only explained by the presence of a polaronic vanadium atom ($V^{2+}$), which is clearly missing otherwise. Our calculations predict that the polaron-induced structural distortion in $VI_3$ is similar to $VCl_3$, i.e. the V-I bond increases by 0.7 Å to screen and accommodate the electron polaron charge. As further supporting evidence of the polaron formation in $VI_3$, the V-2p core levels measured in our $VI_3$ XPS experiments reported in Ref. [28] confirm the presence of both $V^{3+}$ and $V^{2+}$ components, which are compatible with the same polaronic scenario of $VCl_3$. Within the polaronic context, the shoulder observed at the $VI_3$ VB edge [6, 7, 28] is now interpreted as $a_{1g}$ polaronic states and the experiments of Bergner *et al.* [7] find an elegant explanation. Furthermore, the increasing population of the $a_{1g}$ localized V-3d states by electron injection observed by Bergner *et al.* in

VI$_3$ clearly indicates that V atoms continuously trap electrons eventually creating V$^{2+}$ polarons. We verified this last occurrence also for VCl$_3$ simulating the VCl$_3$ DOS with an increasing polaronic concentration (see Fig. S7 [21]) and in fact, we observe a clear enhancement of the V-3d localized state intensity, without any modification of its binding energy.

## III. CONCLUSION

In conclusion, our study provides the first experimental reference for the synthesis and characterization of the single-crystalline VCl$_3$, accompanied by a combined experimental and theoretical understanding of its electronic nature. Photoemission experiments and first-principles DFT calculations clarify the Mott-Hubbard insulating behavior of the material and prove the existence of an extrinsic 2D polaronic phase, which determines the crossing of dispersion-less spin-polarized V-3d $a_{1g}$ states from the conduction to the valence bands with a band inversion with $e'_g$ states. The polaron picture not only well describes quantitatively the VCl$_3$ electronic structure but could offer a new key to explain the complex electromagnetic behavior and spin transport properties of the entire VX$_3$ crystal class, which is still debated. Our findings demonstrated that vanadium trihalides are a genuine case of 2D materials hosting real 2D magnetic polarons, and further enrich the appealing technological applications of this class of materials in the field of spintronics, memory storage, and quantum computing.

## IV. ACKNOWLEDGMENTS

We acknowledge P. Barone and M. Reticcioli for illuminating discussions in the early stage of the research and C. Tresca for his technical support. D. M. acknowledges the CERIC-ERIC Consortium for access to experimental facilities and financial support (proposal grants No. 20222026 and 20227063). G. P. acknowledges support from CINECA Supercomputing Center through the ISCRA project and financial support from the Italian Ministry for Research and Education through the PRIN-2017 project "Tuning and understanding Quantum phases in 2D materials - Quantum 2D" (IT-MIUR Grant No. 2017Z8TS5B). This work has been funded by the European Union - NextGenerationEU under the Italian Ministry of University and Research (MUR) National Innovation Ecosystem grant ECS00000041 - VITALITY - CUP E13C22001060006.

## APPENDIX

### Experimental Methods

High-quality single crystals of VCl$_3$ were obtained through re-crystallization of the commercially available polycrystalline VCl$_3$ (Sigma-Aldrich, 97% purity) through physical vapor transport. The powder was weighed in an argon-filled glove box and then sealed in a silica tube with a diameter D = 10 mm and length L = 180 mm. The ampoule was placed in a horizontal tube furnace in a temperature gradient of 500-310 °C and held there for a week, with the powder positioned in the hot zone. This method yielded flake-like VCl$_3$ crystals of size up to 1.5 × 1.5 × 0.1 mm, grown on top of the polycrystalline powder at the source end of the ampoule. The size of the obtained single crystals was also found to strongly depends on the amount of the starting powder, with the best results obtained for smaller quantities of 200-250 mg. The growth followed the principles of the SSVG technique [18], in which crystals grow on the coolest part of the source powder which acts as a seed. SSVG requires very small temperature gradients; however, our preliminary attempt to recreate the crystal growth in a standard chamber furnace at 500°C, using only the natural horizontal temperature gradient, yielded crystals of unsatisfactory size. The as-grown VCl$_3$ crystals were sealed in silica tubes with a protected Ar atmosphere.

SEM measurements were carried out using a Zeiss Gemini SEM 500. An Oxford Instruments Ultim Max 100 detector was used for the EDX spectroscopy.

Determination of the crystal structure, as well as the orientation of the as-grown VCl$_3$ crystals, were done at room temperature by means of XRD using a Bruker D8 Phaser diffractometer equipped with a Cu K$_\alpha$ radiation source and a LynxEye-XE detector. A single crystal plate was laid down on a zero-background Si holder and the XRD pattern was collected in the 2$\theta$ range from 10° to 120°. The XRD pattern of as-grown VCl$_3$ was processed by means of the LeBail refinement method using the TOPAS software.

The temperature dependence of heat capacity was measured using the Evercool II Quantum Design Physical Property Measurement System (PPMS). The measurements were conducted using the two-$\tau$ relaxation method, in the (1.9 - 300) K temperature range. The crystals were pressed in order to obtain a compact pellet and mounted on a sapphire plate using Apiezon H grease.

For the XPS investigation, VCl$_3$ crystals were mechanically exfoliated in a protected N$_2$ environment using the tape method [48]. The last piece of tape, loaded with thin VCl$_3$ flakes, was attached directly to carbon tape, which was used as sticky conductive substrate, and then inserted into the ultra-high vacuum XPS setup. Finally, the tape was peeled off *in-situ* to have VCl$_3$ flakes with clean surfaces. XPS analysis was performed using a monochromatic Al-K$_\alpha$ source (h$\nu$ = 1486.6 eV) and a multichannel electron energy analyzer (Specs Phoibos 150). XPS data were recorded at room temperature. During preliminary XPS measurements, thick VCl$_3$ flakes exhibit massive charging effects upon X-ray flux exposure. To avoid charging of VCl$_3$, the thick crystals were mechanically exfoliated many times to thin enough the VCl$_3$ flakes. The analysis of the core levels and valence bands was carried out after Shirley background subtraction and calibration to the C-1s core level binding energy set at a binding energy of 284.6 eV (C=C bond). Voigt functions were used to decompose the XPS core levels spectra.

### Computational Methods

DFT calculations were performed using the Vienna *ab initio* Simulation Package (VASP) [49, 50], within the generalized gradient approximation (GGA), and the Perdew-Burke-Ernzerhof (PBE) exchange-correlation functional [51]. Interactions between electrons and nuclei were described using the projector-augmented wave (PAW) method. Energy and force thresholds were set to $10^{-5}$ eV and $10^{-4}$ eVÅ$^{-1}$, respectively. A plane-wave kinetic energy cutoff of 450 eV was employed. The Brillouin zone (BZ) was sampled using $7\times7\times7$ Gamma-centered Monkhorst-Pack grids for the bulk phases, while a $12\times12\times1$ grid was used for the monolayer phase. To account for the on-site electron-electron correlation we used the GGA+U approach with an effective Hubbard term $U$=3.5 eV consistent with the value calculated by He *et al.* [52] with linear response theory [53]. The van der Waals interactions between the VCl$_3$ layers were considered by adopting the DFT-D2 method [54]. The calculations for the bulk phase were carried out using the relaxed lattice parameters ($a$=6.084 Å and $c$=17.625 Å) with the spin-polarized PBE+U+vdW method (see Tab. S1 [21]). The monolayer phase was described in the supercell approach using a relaxed lattice parameter of $a$=6.084 Å and a vacuum region of 15 Å.

The same approach was used for the DOS of VI$_3$ shown in Fig. 5. To account for the on-site electron-electron correlation we used the GGA+U approach with an effective Hubbard term $U$=3.7 eV consistent with the value calculated by He *et al.* [52] with linear response theory [53]. The experimental bulk VI$_3$ lattice parameter was used in the calculations ( $a$= 6.93 Å [55]).

Supplemental material for

# Polaronic and Mott insulating phase of layered magnetic vanadium trihalide VCl$_3$

Dario Mastrippolito,[a,*] Luigi Camerano,[a] Hanna Światek,[b,c] Břetislav Šmíd,[d] Tomasz Klimczuk,[b,c] Luca Ottaviano,[a,e] Gianni Profeta[a,e]

[a] Department of Physical and Chemical Sciences, University of L'Aquila, Via Vetoio 67100 L'Aquila, Italy
[b] Faculty of Applied Physics and Mathematics Gdansk University of Technology, Gdansk, Poland
[c] Advanced Materials Center, Gdansk University of Technology, Gdansk, Poland
[d] Charles University, Faculty of Mathematics and Physics, Department of Surface and Plasma Science, V Holešovičkách 2, 180 00 Prague 8, Czech Republic
[e] CNR-SPIN L'Aquila, Via Vetoio 67100 L'Aquila, Italy

[*] E-mail: dario.mastrippolito@graduate.univaq.it

**CONTENTS**

## I. CRYSTAL CHARACTERIZATION

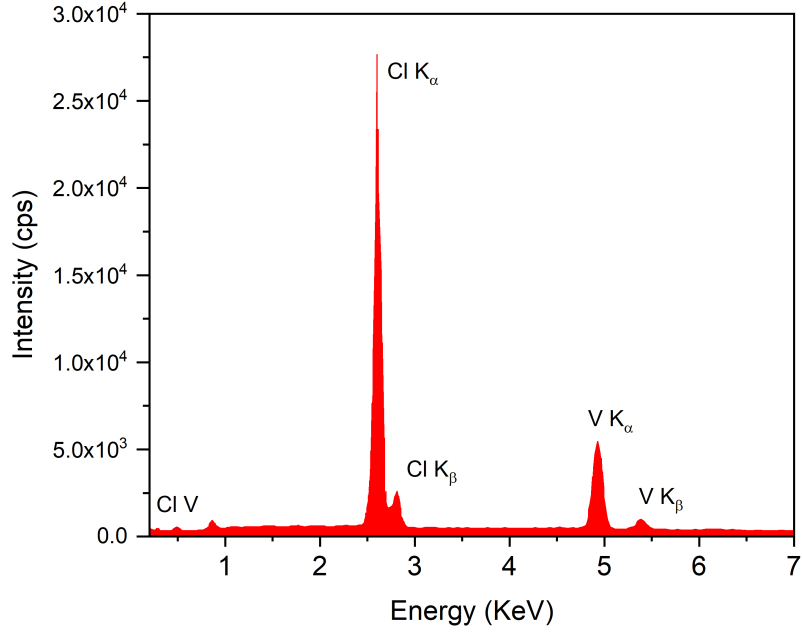

FIG. S1. EDX micro-analysis of as-grown VCl$_3$ crystals.

## II. OPTIMIZED LATTICE PARAMETERS

TABLE S1. VCl$_3$ lattice parameters: comparison between calculated and measured.

|  | $a$ (Å) | $c$ (Å) |
|---|---|---|
| PBE+U+vdW | 5.798 | 17.881 |
| PBE+U (spin polarized) | 6.133 | 18.453 |
| PBE+U+vdW (spin polarized) | 6.084 | 17.625 |
| experiment |  | 17.49 |

## III. DENSITY OF STATES FOR MONOLAYER AND BULK PHASES

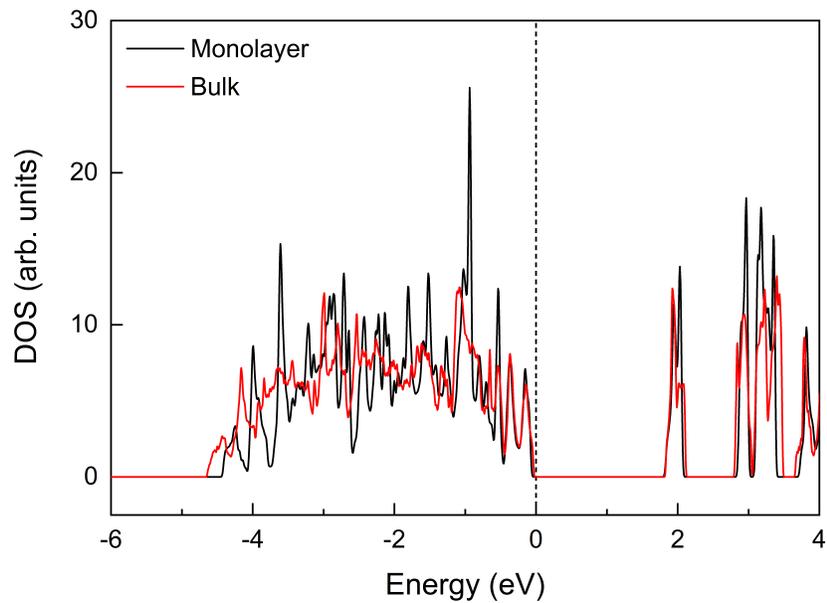

FIG. S2. Density of states of VCl$_3$ in the monolayer (black) and bulk (red) phases.

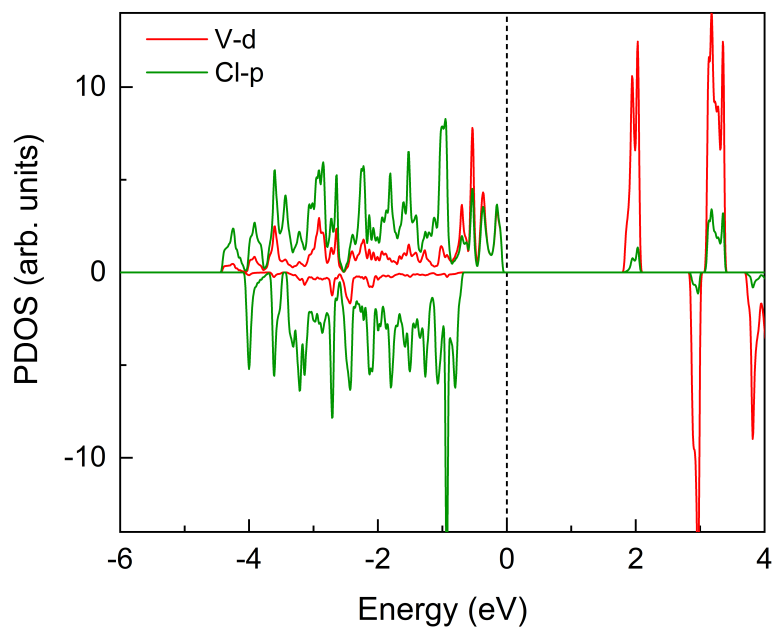

FIG. S3. Spin-resolved partial density of states of VCl$_3$ monolayer.

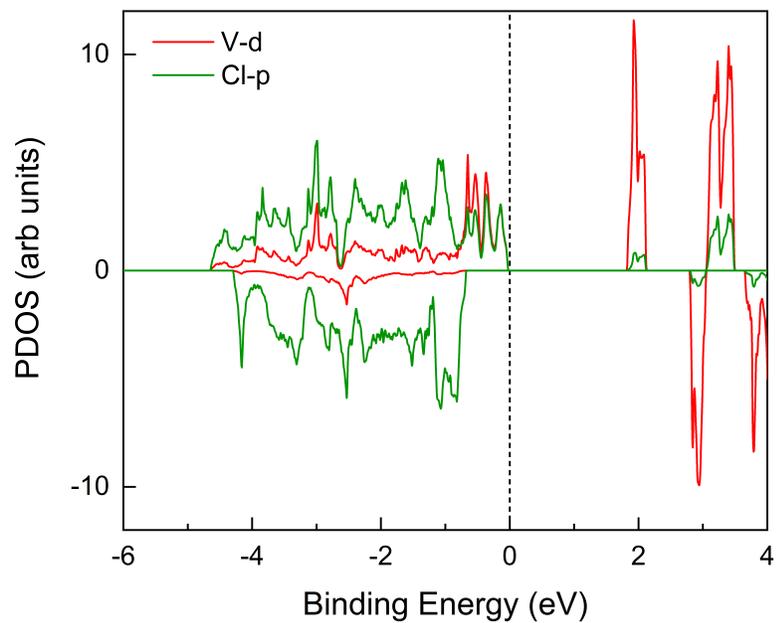

FIG. S4. Spin-resolved partial density of states of bulk $VCl_3$.

## IV. DENSITY OF STATES WITH SPIN-ORBIT COUPLING

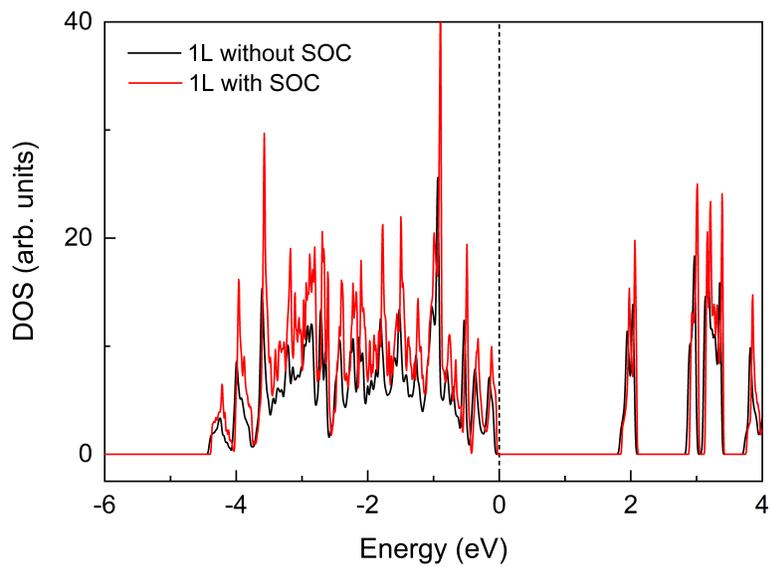

FIG. S5. Density of states of VCl$_3$ monolayer (1L) with (red) and without (black) spin-orbit coupling.

## V. ELECTRON-DOPED AND POLARON CONFIGURATIONS

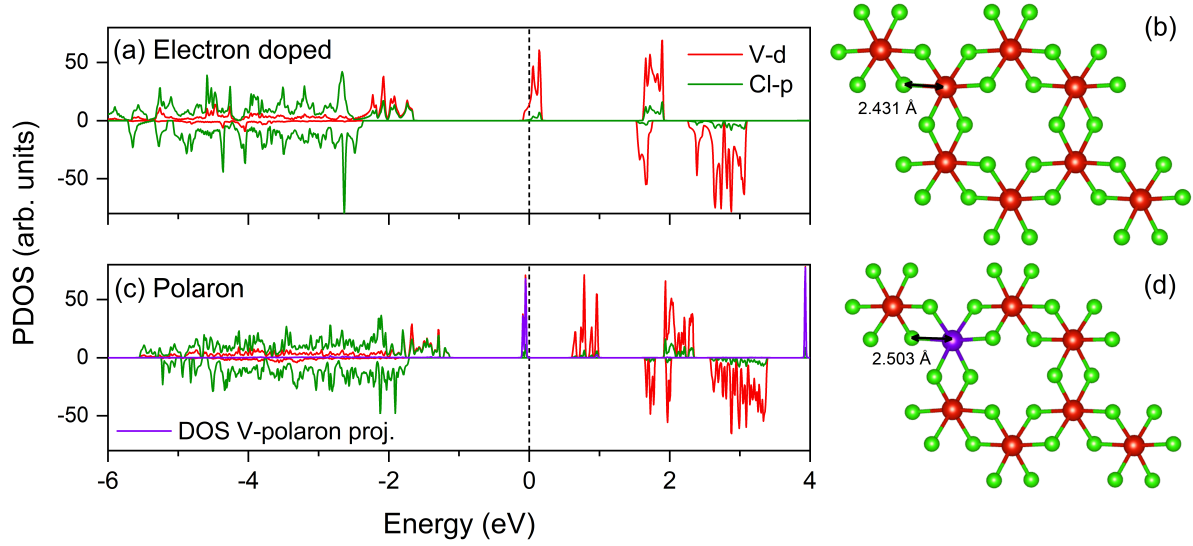

FIG. S6. Spin-resolved partial density of states (panel (a)) and distorted crystal structure (panel (b)) of the jellium doped 2×2 VCl$_3$ supercells. Spin-resolved density of states (panel (c)) and distorted crystal structure of the polaronic phase of VCl$_3$ (panel (d)). V and Cl atoms are represented in red and green spheres, respectively. The violet atom represents the V$^{2+}$ polaronic atom.

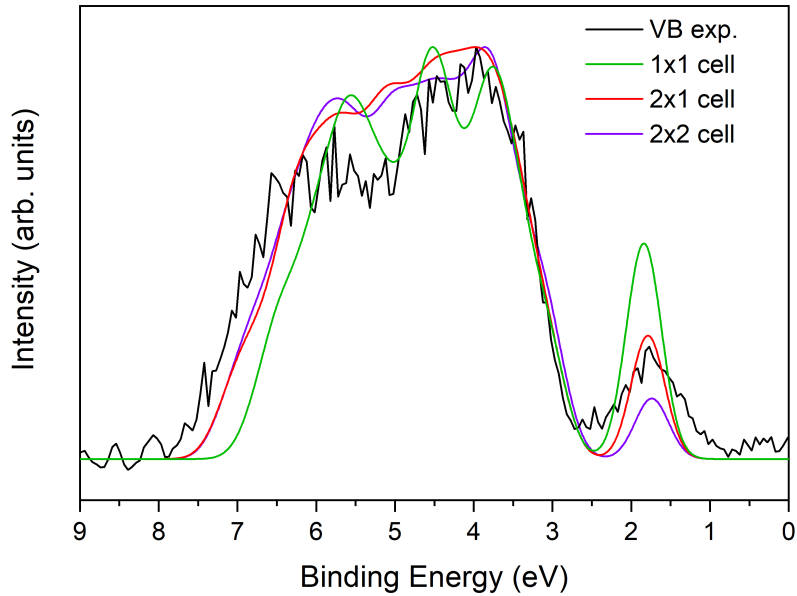

FIG. S7. Experimental valence bands spectrum of VCl$_3$ (black curve) compared with the density of states of VCl$_3$ in the polaronic phase, calculated using different cells: 1×1 (green, 0.8 % electron doping), 2×1 (red, 0.4 % electron doping), and 2×2 (violet, 0.2 % electron doping). The densities of states are convoluted with a Gaussian function ($\sigma = 0.5$ eV) to account for the experimental broadening.

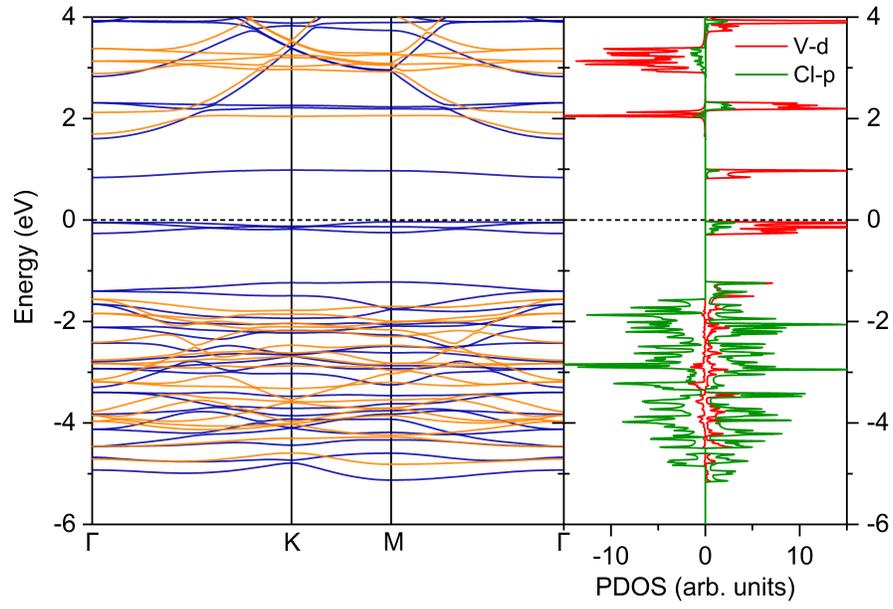

FIG. S8. Spin-resolved band structure (left panel: blue and orange represent spin up and spin down components, respectively) and partial density of states of unit cell of VCl$_3$ in the polaronic phase.